# Sm-Co-based amorphous alloy films for zero-field operation of transverse thermoelectric generation

Rajkumar Modak[a], Yuya Sakuraba[a], Takamasa Hirai[a], Takashi Yagi[b], Hossein Sepehri-Amin[a], Weinan Zhou[a], Hiroto Masuda[c], Takeshi Seki[a,c,d], Koki Takanashi[c,d,e,\*], Tadakatsu Ohkubo[a], and Ken-ichi Uchida[a,c,d]

[a]National Institute for Materials Science, Tsukuba, 305-0047, Japan

[b]National Institute of Advanced Industrial Science and Technology, Tsukuba 305-8565, Japan

[c]Institute for Materials Research, Tohoku University, Sendai 980-8577, Japan

[d]Center for Spintronics Research Network, Tohoku University, Sendai 980-8577, Japan

[e]Center for Science and Innovation in Spintronics, Tohoku University, Sendai 980-8577, Japan

*Present address: Advanced Science Research Center, Japan Atomic Energy Agency, Tokai, Ibaraki 319-1195, Japan

Corresponding authors: MODAK.Rajkumar@nims.go.jp (Rajkumar Modak), UCHIDA.Kenichi@nims.go.jp (Ken-ichi Uchida)

**ABSTRACT**

Transverse thermoelectric generation using magnetic materials is essential to develop active thermal engineering technologies, for which the improvements of not only the thermoelectric output but also applicability and versatility are required. In this study, using combinatorial material science and lock-in thermography technique, we have systematically investigated the transverse thermoelectric performance of Sm-Co-based alloy films. The high-throughput material investigation revealed the best Sm-Co-based alloys with the large anomalous Nernst effect (ANE) as well as the anomalous Ettingshausen effect (AEE). In addition to ANE/AEE, we discovered unique and superior material properties in these alloys: the amorphous structure, low thermal conductivity, and large in-plane coercivity and remanent magnetization. These properties make it advantageous over conventional materials to realize heat flux sensing applications based on ANE, as our Sm-Co-based films can generate thermoelectric output without an external magnetic field. Importantly, the amorphous nature enables the fabrication of these films on various substrates including flexible sheets, making the large-scale and low-cost manufacturing easier. Our demonstration will provide a pathway to develop flexible transverse thermoelectric devices for smart thermal management.

## 1. Introduction

In recent years, the research on spin caloritronics is showing a new direction with the demonstration of many magneto-thermoelectric effects by advanced heat detection techniques [1–5]. The magneto-thermoelectric effects that provide interconversion between heat and charge currents in magnetic materials exhibit unique thermoelectric conversion functionalities, which potentially enable versatile thermal energy harvesting and active thermal management for electronic and spintronic devices. Among the magneto-thermoelectric effects, the anomalous Nernst effect (ANE) and its Onsager reciprocal called anomalous Ettingshausen effect (AEE) are of particular interest due to their unique heat-charge current conversion symmetry [6–12]. ANE (AEE) generates a charge (heat) current in the direction perpendicular to an applied temperature gradient $\nabla T$ (charge current) and magnetization **M**:

$$\mathbf{j}_{c,\text{ANE}} = \sigma S_{\text{ANE}}(\hat{\mathbf{m}} \times \nabla T), \quad (1)$$

$$\mathbf{j}_{q,\text{AEE}} = \Pi_{\text{AEE}}(\hat{\mathbf{m}} \times \mathbf{j}_c), \quad (2)$$

where $\mathbf{j}_{c,\text{ANE}}$ ($\mathbf{j}_{q,\text{AEE}}$) is the charge (heat) current density driven by ANE (AEE), $\mathbf{j}_c$ is the charge current density applied to a magnetic material, $\hat{\mathbf{m}}$ is the unit vector along **M**, $S_{\text{ANE}}$ ($\Pi_{\text{AEE}}$) is the ANE (AEE) coefficient, and $\sigma$ is the electrical conductivity. The above relations provide various advantages for ANE/AEE-based devices. For example, the symmetry of ANE enables thermoelectric generation simply by forming a magnetic material on a heat source surface. This simple and effective structure makes it possible to construct thermoelectric modules with high flexibility, high mechanical endurance, and low cost and to reduce efficiency losses in contact resistances [13]. The output voltage and power induced by ANE can be increased by increasing the length and area of the magnetic material in a lateral direction, respectively. Since it is easy to increase the length and area of such lateral devices, ANE is suitable for harvesting thermal energy distributed over a large area. In contrast, AEE enables active control of heat current directions and distributions by controlling the **M** direction [14]. These strong potentials of ANE/AEE have stimulated innumerable studies in fundamental and materials science in these effects.

Despite these technological advantages, practical applications of ANE/AEE-based thermoelectric converters require systematic device engineering and introduction of new materials as most of the conventional ANE/AEE materials till date have low thermoelectric conversion performance. To overcome this situation, researchers around the globe are investigating ANE/AEE systematically and many ANE/AEE materials have been discovered. Recently, large $S_{ANE}$ of 6 ~ 8 ×10$^{-6}$ VK$^{-1}$ was observed in ferromagnetic Heusler compound Co$_2$MnGa, which is the record-high value at room temperature [15–18]. In 2021, Asaba *et al*. found giant $S_{ANE}$ of 23 ×10$^{-6}$ VK$^{-1}$ at 40 K in UCo$_{0.8}$Ru$_{0.2}$Al [19]. Large ANE was also observed in single-crystalline noncollinear antiferromagnet YbMnBi$_2$, where $S_{ANE}$ ~ 6 ×10$^{-6}$ VK$^{-1}$ at 160 K [20]. However, most of ANE/AEE materials require an application of external magnetic fields to align the magnetization; this situation significantly compensates the advantages of ANE/AEE-based devices and limits the practical applications. Therefore, ANE/AEE materials with large remanent magnetization and coercive force are desired. In this context, Miura *et al*. found that SmCo$_5$-type bulk permanent magnets exhibit large AEE and their figure of merit above room temperature is comparable to that in Co$_2$MnGa [8,11]. This finding suggests that the investigation of ANE/AEE in rare-earth magnets is important to realize zero-field operation of transverse-thermoelectric conversion. Although bulk ANE/AEE materials are suitable for energy harvesting and temperature control, thin films are advantageous for smart applications including wearable thermoelectric convertors and heat flux sensors due to high flexibility and low heat resistance.

In this paper, we show that Sm-Co-based amorphous thin films enable zero-field operation of ANE/AEE. Using combinatorial thin-film deposition and thermoelectric imaging techniques, we systematically investigated the composition dependence of AEE for Sm$_p$Co$_{100-p}$ ($0 \leq p \leq 100$) amorphous films and confirmed that Sm$_{20}$Co$_{80}$ shows largest AEE-induced temperature modulation. We demonstrated that AEE can further be improved by partially replacing Co with Fe by preparing Sm$_{20}$(Co$_{100-q}$Fe$_q$)$_{80}$ ($0 \leq q \leq 100$) composition-spread films and found that Sm$_{20}$(Co$_{77}$Fe$_{23}$)$_{80}$ exhibits largest AEE in these alloys. Despite their amorphous nature, the Sm-Co-based films exhibit large in-plane remanent magnetization and coercive force; these films are advantageous over conventional materials for applications including heat flux sensors, as they can generate thermoelectric output in the absence of external magnetic fields. Importantly, the Sm-Co-based amorphous films can be fabricated on any surfaces including flexible substrates, which improves the versatility of the ANE/AEE devices.

## 2. Experimental details

### *2.1. Sample preparation*

The $Sm_pCo_{100-p}$ and $Sm_{20}(Co_{100-q}Fe_q)_{80}$ composition-spread films with a thickness of 100 nm were fabricated on a single-crystalline MgO (001) substrate using a magnetron sputtering system (Comet, Inc., CMS-320), where direct current power sources were used for Sm and Co depositions and a radio frequency power source was used for Fe deposition. Before sputtering the films, the chamber was evacuated at a base pressure of $< 6.0 \times 10^{-6}$ Pa and then deposited at a process Ar gas pressure of 0.4 Pa at ambient temperature. To achieve a composition variation of $0 \leq p \leq 100$ at.% over a length of 7.0 mm on a single substrate, the films were prepared from Sm and Co targets using a layer-by-layer wedge shaped deposition process established in earlier studies [21,22]. Here, (I) we first deposited a wedge-shaped Co layer using a linear moving shutter over a length of 7.0 mm, (II) then rotated the substrate by 180º, (III) followed by the deposition of a wedge-shaped Sm layer using the linear moving shutter over a length of 7.0 mm, where the total thickness of the $Sm_pCo_{100-p}$ films after completing (I)-(III) were designed to be 1.0 nm. This sequence was repeated 100 times to get the 100-nm-thick composition-spread film (see Figure 1(a)). The $Sm_{20}(Co_{100-q}Fe_q)_{80}$ composition-spread films with *q* varying from 0 to 100 at.% were prepared using the similar approach, where first a uniform layer of Sm was deposited and then wedge shaped Co and Fe layers with an opposite thickness gradient were deposited on top of the Sm layer (see Figure 7(a)). The combination of the uniform Sm layer and the Co/Fe wedge layer forms the 1-nm-thick $Sm_{20}(Co_{100-q}Fe_q)_{80}$ film. The processes were repeated 100 times to get the 100-nm-thick film. In addition to the composition-spread films, we also fabricated uniform 100-nm-thick $Sm_{20}Co_{80}$ and $Sm_{20}(Co_{77}Fe_{23})_{80}$ films by the repeating Sm/Co and Sm/Co/Fe layers for 100 times, respectively (see Figure 5(a) and Figure 9(a)). All the films were capped with a 2-nm-thick Al film to prevent oxidation.

## 2.2. Composition, structural, and magnetic characterization

The Sm, Co, and Fe concentrations of the $Sm_pCo_{100-p}$ and $Sm_{20}(Co_{100-q}Fe_q)_{80}$ composition-spread films were measured by the thermal emission electron probe microanalyzer (EPMA) (JEOL, JXA-8530F). The crystal structures of the composition-spread films were investigated by the X-ray diffraction (XRD) method (Rigaku, SmartLab) with a Cu $K_\alpha$ X-ray source ($\lambda = 0.15406$ nm). The X-ray was incident on the sample in the Bragg-Brentano geometry as a parallel beam through a length limit slit of 0.5 mm in the direction perpendicular to the composition gradient and detected with a 2D detector (Rigaku, PILATUS100K/R). The measurements were performed at different positions on the $Sm_pCo_{100-p}$ and $Sm_{20}(Co_{100-q}Fe_q)_{80}$ composition-spread films along the composition gradient at an interval of 1.0 mm. The microstructures for the films were investigated using scanning transmission electron microscope (STEM) (FEI, Titan G2 80-200) equipped with a probe aberration corrector. energy-dispersive-spectroscopy (EDS) data was collected using a Super-X EDX detector. The STEM specimens were prepared by a FEI Helios G4 dual beam machine using the lift-out process. The composition dependence of the magnetic property of the $Sm_pCo_{100-p}$ and $Sm_{20}(Co_{100-q}Fe_q)_{80}$ composition-spread films was evaluated by performing position-dependent in-plane micro-magneto-optical Kerr effect (MOKE) (NEOARK Co., BH-PI920BU-TKT) measurement at an interval of 0.25 mm along the composition gradient direction using a green laser with spot size of 2 μm diameter in minimum. The magnetization curve of the uniform $Sm_{20}Co_{80}$ and $Sm_{20}(Co_{77}Fe_{23})_{80}$ films were characterized using the vibrating sample magnetometer (Quantum Design, MPMS3) at room temperature.

### *2.3. Transport properties measurement*

The composition dependence of the electrical resistivity ($\rho_{xx}$) for the $Sm_{20}(Co_{100-q}Fe_q)_{80}$ composition-spread film was measured by means of the four-probe method, where the distance between the voltage probes is 100 μm and the local resistivity measurements were performed at an interval of 0.25 mm. The $\rho_{xx}$ and Seebeck coefficient $S_{SE}$ values for the uniform $Sm_{20}Co_{80}$ and $Sm_{20}(Co_{77}Fe_{23})_{80}$ films were measured using the Seebeck Coefficient / Electric Resistance Measurement System (ADVANCE RIKO, ZEM-3). To measure ANE and AHE for the uniform $Sm_{20}Co_{80}$ and $Sm_{20}(Co_{77}Fe_{23})_{80}$ films, Hall-bar-shaped patterns were prepared using photolithography and Ar-ion milling. The ANE voltage was measured by applying a temperature gradient and magnetic field in the in-plane and out-of-plane directions, respectively (see the inset to Figure 5(c)). The temperature distribution was precisely evaluated by an infrared camera while coating the sample surface with the black ink to correct an emissivity, which is a technique established in earlier studies [17,23–25]. To evaluate AHE, the Hall voltage along the width direction of the Hall bar was measured by applying a charge current of 1 mA in the length direction of the sample and the external magnetic field in the out-of-plane direction.

### *2.4. Lock-in thermography (LIT) measurement*

For the AEE measurement using the LIT technique, the $Sm_pCo_{100-p}$ and $Sm_{20}(Co_{100-q}Fe_q)_{80}$ composition-spread films were patterned into a device consisting of two wires with a width of 0.4 mm and length of 8.0 mm, where the composition gradient was along the length direction. During the LIT measurements, one ends of the wires were electrically connected, such that the two wires were connected in series with the charge current flowing along the opposite directions (see Figure 4(a)). The magnetic field was applied along the $y$ direction (width direction of the wires); thus, the temperature modulation is generated on the sample surface due to the AEE-induced heat current along the thickness direction. The infrared radiation thermally emitted from the sample surface was detected while applying a square-wave-modulated AC charge current with the square-wave amplitude of 15 mA, frequency $f =$ 25 Hz, and zero DC offset along the wires and an external magnetic field with the magnitude $\mu_0 H = \pm 0.8$ T. To extract the pure AEE contribution from other thermoelectric background signals, we employed the previously established procedure [6–10,26,27]; we extracted lock-in amplitude $A_{\text{odd}}$ and phase $\phi_{\text{odd}}$ images showing the distribution of the current-induced temperature modulation with the $H$-odd dependence, which were obtained by subtracting the raw LIT images for $\mathbf{M} \parallel +y$ from those for $\mathbf{M} \parallel -y$ and dividing the subtracted images by 2.

## 2.5. Time-domain thermoreflectance (TDTR) measurement

The out-of-plane thermal conductivity for the $Sm_{20}(Co_{100-q}Fe_q)_{80}$ composition-spread film was measured by means of a pump-probe optical technique called TDTR [28,29]. For the TDTR measurements, the $Sm_{20}(Co_{100-q}Fe_q)_{80}$ composition-spread film was covered with a 100-nm-thick Al capping layer. The Al layer works as a transducer because the thermoreflectance coefficient of Al is well known [29–31]. The TDTR system used in this study is based on the front-heating front-detection configuration [32], where the pulse duration, optical power, repetition rate, modulation frequency, wavelength, and spot diameters of the pump (probe) laser are 0.5 ps (0.5 ps), 47 mW (< 1 mW), 20 MHz (20 MHz), 200 kHz (none), 1550 nm (775 nm), and 90 μm (~30 μm), respectively. Since the diameter of the pump laser spot is much larger than the heat diffusion length in the MgO substrate, $\sqrt{\alpha/\pi f} \cong 5.5$ μm, where $\alpha$ is the thermal diffusivity of MgO and $f$ is the modulation frequency, one-dimensional heat diffusion can be presumed. The pump laser was irradiated on the film surface, and then the probe laser was focused at the same spot with the specific delay that is electrically controlled by a function generator [32]. By detecting the reflected probe laser using a Si adjustable balanced photoreceiver connected to a lock-in amplifier, the amplitude and phase components of thermoreflectance signals synchronized with the modulation frequency can be characterized. Finally, the thermal conductivity of the $Sm_{20}(Co_{100-q}Fe_q)_{80}$ layer was quantified from the measured TDTR signals with a heat conduction model [32], where a one-dimensional heat conduction in the model comprising the MgO substrate and two layers, i.e., $Sm_{20}(Co_{100-q}Fe_q)_{80}$ and Al, are assumed. Here, the actual thicknesses of the $Sm_{20}(Co_{100-q}Fe_q)_{80}$ and Al layers are estimated based on cross-sectional STEM measurements.

## 3. Results and discussion

### *3.1. Deposition and characterization of $Sm_pCo_{100-p}$ ($0 \leq p \leq 100$) composition-spread film*

The $Sm_pCo_{100-p}$ composition-spread film was prepared on a MgO (001) substrate using magnetron sputtering, where the Sm concentration ($p$) is varied from 0 at.% (pure Co) to 100 at.% (pure Sm) on a single film. The wedge-shaped Sm and Co layers were alternately deposited using a linear moving shutter and the thickness gradient direction of the Sm layers was kept opposite to that of the Co layers. The thickest part of each layer has a thickness of 1 nm and one Sm-Co pair forms a uniform 1-nm-thick film, where the Sm and Co atoms are naturally mixed without any heat treatment as a solid solution or binary alloy. By repeating the deposition of such Sm-Co pairs 100 times, a 100-nm-thick $Sm_pCo_{100-p}$ composition-spread film was obtained (see experimental section for detailed procedures). The systematic change of the $Sm_pCo_{100-p}$ alloy composition along the composition gradient was confirmed by position-dependent EPMA measurements. A schematic illustration of the deposited film is displayed in Figure 1(a). Such a composition-spread film is useful for high-throughput material screening over wide composition range without time-consuming experiments using many films.

Figures 1(b-c) respectively represent the XRD line profile and 2D image profile of the as-deposited $Sm_pCo_{100-p}$ composition-spread film measured at different positions along the composition gradient. The XRD data reveals that the $Sm_pCo_{100-p}$ alloys are amorphous in nature except the Co-rich and Sm-rich regions. For the Co-rich and Sm-rich regions, the film has polycrystalline nature, where the pure metal phases are dominated and hcp Co and trigonal Sm are formed (see the 2D XRD profiles at $x$ = 1.5 mm, 2.5 mm, and 8.5 mm).

Figures 2(a-c) show the cross-sectional bright-field (BF) STEM images with microbeam electron diffraction patterns. The electron diffraction patterns clearly demonstrate no crystalline ordering at the $Sm_{30}Co_{70}$ composition, while the crystalline ordering was observed for the Co-rich and Sm-rich regions. These results are consistent with the XRD data. Figure 2(d) show the STEM-EDS elemental maps obtained at Co-rich, $Sm_{30}Co_{70}$, and Sm-rich regions of the $Sm_pCo_{100-p}$ composition-spread film. Figures 2(e-f) display the high-magnification STEM-EDS maps of Sm and Co observed from the $Sm_{30}Co_{70}$ region of the film and composition line profiles of the constituent elements (obtained from selected box shown in Figure 2(e)), respectively. These data reveal the formation of the $Sm_pCo_{100-p}$ alloy with mixing of Sm and Co atoms in microstructure level, although the film was fabricated by layer-by-layer deposition.

To investigate the magnetic properties of this amorphous $Sm_pCo_{100-p}$ alloys, we measured the composition-dependent MOKE loops using the above composition-spread film with applying an in-plane magnetic field. From the hysteresis behavior of the MOKE loops, we estimated the composition dependence of the coercive field ($\mu_0 H_C^{\mathrm{Kerr}}$) and plotted as a function of $p$ in Figure 3(a). We observed that the alloy compositions with $20 \leq p \leq 30$ at.% have significantly large in-plane coercivity and remanent magnetization. The typical Kerr loops observed at $p$ = 21 at.% and 43 at.% for the film are shown in Figures 3(b-c), respectively. The absence of the MOKE signal for $p >$ 43 at.% signifies that the $Sm_pCo_{100-p}$ alloys with $p >$ 43 at.% are nonmagnetic (see the inset to Figure 3(a)). This magnetic behavior is consistent with previous works on Sm-Co based amorphous films [33–35]. The magnetic anisotropies originated from short range ordering of SmCo alloys in these amorphous structures are considered to be the origin of such magnetic nature. However, the detailed microscopic origin for this property remains to be investigated. The presence of the large in-plane coercivity and remanent magnetization make these alloys potential candidates for the zero-field operation of the transverse thermoelectric conversion.

### 3.2. Transverse thermoelectric conversion performance of $Sm_pCo_{100-p}$ films

Now we are in a position to evaluate the composition dependence of the transverse thermoelectric effects, i.e., ANE and AEE, in the $Sm_pCo_{100-p}$ alloys. The direct one-by-one measurements of ANE and other transport properties using conventional Hall-bar-shaped samples for a composition-spread film involve complicated device design and time-consuming experiments. Also, the limitation of the number of the Hall bars along the composition gradient lowers the composition resolution. To solve such hurdles, we employed the LIT technique which enables imaging measurements of the temperature modulations due to thermoelectric effects [1–3,36,37]. In the LIT measurements, the thermal images recorded with applying a charge current to a sample are Fourier transformed to obtain the lock-in amplitude $A$ and phase $\phi$ images, where $A$ and $\phi$ represent the magnitude and sign of temperature modulation due to the thermoelectric effects. As can be seen from Equation (2) and Figure 1(a), if a charge current is applied along the $x$ direction (along the composition gradient) and magnetic field is applied along the $y$ direction, the AEE-induced heat current is generated along the $z$ direction (perpendicular to the film plane). Thus, the LIT image recorded in this configuration contains the spatial distribution of temperature modulation signals reflecting the composition dependence of AEE. Therefore, the AEE measurements using LIT require only a simple strip along the composition gradient, thus enabling easy and efficient scanning to find the suitable composition region with largest AEE in a composition-spread film. Importantly, the magnitude of the AEE-induced temperature modulation obtained by extracting the line profile of the LIT image is continuous along the composition gradient. This gives much accurate estimation of the composition dependence of AEE [6]. Based on the reciprocal relation between AEE and ANE, the combination of composition-spread film and LIT enables high-throughput screening of ANE materials.

Figures 4(a-d) represent the lock-in $A$ and $\phi$ images for the Sm$_p$Co$_{100-p}$ composition-spread film at the charge current amplitude $J_c$ = 15 mA and frequency $f$ = 25 Hz for various magnitudes of an external magnetic field $\mu_0 H$. The images were recorded while applying the magnetic field in the order of $\mu_0 H$ = 0.8 T (Figure 4(a)), 0.0 T (Figure 4(b)), -0.8 T (Figure 4(c)), and 0.0 T (Figure 4(d)) (note that our MOKE measurements confirm that the in-plane magnetization saturates at $|\mu_0 H|$ = 0.8 T for any compositions). Thus, by subtracting the LIT images in Figure 4(c) (Figure 4(d)) from those in Figure 4(a) (Figure 4(b)), we can obtain the $A_{\text{AEE}}$ and $\phi_{\text{AEE}}$ images showing the pure AEE-induced temperature modulation for the saturation magnetization (remanent magnetization) state, as shown in Figure 4(e) (Figure 4(f)). The $\phi_{\text{AEE}}$ image in Figure 4(e) shows that the sign of the temperature modulation for the top wire is opposite to that of the bottom wire because of the phase difference of ~180˚. The result indicates that $\mathbf{j}_{q,\text{AEE}}$ is generated along the +$z$ (-$z$) direction when $\mathbf{M}$ and $\mathbf{J}_c$ are along the +$y$ and -$x$ (+$x$) directions, respectively, which is consistent with the symmetry of AEE in Equation (2). Figure 4(f) shows that similar $A_{\text{AEE}}$ and $\phi_{\text{AEE}}$ signals appear in the Sm$_p$Co$_{100-p}$ composition-spread film even in the remanent magnetization state, confirming the zero-field operation of AEE in our film.

Figure 4(g) shows the $x$-directional line profile of $A_{\text{AEE}}$ normalized by the charge current density $j_\text{c}$ for the $\text{Sm}_p\text{Co}_{100-p}$ composition-spread film both in the saturation magnetization ($|\mu_0 H| = 0.8$ T) and remanent magnetization ($|\mu_0 H| = 0.0$ T) states, extracted from the images in Figures 4(e-f), respectively. Here, the position dependence of $A_{\text{AEE}}/j_\text{c}$ is converted into the $p$ dependence by using the $x$ dependence of the alloy composition. We found that the AEE-induced temperature modulation is very small for pure Co and gradually increases with increasing $p$ in the $\text{Sm}_p\text{Co}_{1-p}$ alloy for the $0 \leq p \leq 20$ at.% region. At $p = 20$ at.%, the alloy shows the maximum AEE signal which remains nearly constant for $20 \leq p \leq 30$ at.%. Then, the magnitude of the AEE signal gradually decreases for $30 \leq p \leq 43$ at.% and vanishes for $p > 43$ at.%. The similar tendency was also observed in the remanent magnetization state. Here, it should be noted that the maximum value of $A_{\text{AEE}}/j$ in the remanent magnetization state is nearly 30% reduced from that in the saturation magnetization state. This behavior can be explained by the fact that the magnetic field dependence of the AEE signal follows the magnetization curve (see the MOKE signal in Figure 3(b)).

### 3.3. Measurement of anomalous Nernst coefficient for $\text{Sm}_{20}\text{Co}_{80}$ film

Through the LIT measurements using the $\text{Sm}_p\text{Co}_{100-p}$ composition-spread films, we have successfully determined the compositions showing largest AEE among the present amorphous alloys. It is also important to quantitatively determine $S_{\text{ANE}}$ and $\Pi_{\text{AEE}}$ to evaluate the transverse thermoelectric performance of the amorphous $\text{Sm}_p\text{Co}_{100-p}$ alloys. For this purpose, we fabricated a uniform $\text{Sm}_{20}\text{Co}_{80}$ film with the optimized composition by alternately depositing Sm and Co thin layers (see Figure 5(a)). The thicknesses of each layer were determined from the layer thicknesses of the composition-spread film used for the aforementioned experiment. The $\text{Sm}_{20}\text{Co}_{80}$ film is amorphous in nature as confirmed by a microbeam electron diffraction pattern (see supplementary material Figure S1), which is consistent with the structural property of the composition-spread film at this composition. To investigate the magnetic property of the uniform film, we measured in-plane magnetization curve. As represented in Figure 5(b), the film exhibits hysteresis behaviors with large in-plane coercivity and remanent magnetization. These results suggest that the uniform $Sm_{20}Co_{80}$ film has structural and magnetic behaviors similar to the composition-spread film at the same composition region.

To determine $S_{ANE}$, the uniform $Sm_{20}Co_{80}$ film was patterned into a Hall-bar structure (see the inset to Figure 5(c) and experimental section). ANE was measured with applying the magnetic field along the film thickness direction because the estimation of $S_{ANE}$ requires the accurate determination of the applied temperature gradient $\nabla T$ and thus $\nabla T$ must be applied along the in-plane direction [12]. Figure 5(c) shows the out-of-plane $H$ dependence of the transverse voltage $V$ normalized by the Hall bar width $w$ for various values of $\nabla T$. The electric field generated by ANE $E_{ANE}$ (=$V_{ANE}/w$) was extracted from the field-odd-dependent component of $V$ at the field magnitude of $\mu_0 H = 5.0$ T. The $S_{ANE}$ value was evaluated by linear fitting the $\nabla T$ dependence of $E_{ANE}$, as shown in Figure 5(d). The $S_{ANE}$ value for the uniform $Sm_{20}Co_{80}$ film was estimated to be $1.07 \pm 0.08 \times 10^{-6}$ VK$^{-1}$. Although this $S_{ANE}$ value is three times smaller than $S_{ANE}$ for $SmCo_5$-type bulk permanent magnets [8,11], it is much larger than that for pure Co ($S_{ANE} = 0.32 \times 10^{-6}$ VK$^{-1}$)[12,38] and comparable to that for $Sm_{12}Co_{17}$-type and $Nd_2Fe_{14}B$-type bulk magnets [8]. Despite the moderate $S_{ANE}$, our demonstration shown in the next section reveals the usefulness of our amorphous films for ANE-based applications.

### *3.4. Demonstration of heat flux sensor using $Sm_{20}Co_{80}$ film*

Here, we demonstrate the performance of the SmCo-based amorphous films as a flexible heat flux sensor based on ANE. The heat flux sensor is a device enabling quick and simultaneous detection of the magnitude and direction of a heat flow, which can be an essential component for smart thermal management systems. However, till now, the commercial applications of heat flux sensors are confined due to several limitations. The commercially available sensors are based on the Seebeck effect and constructed by a serially connected three-dimensional matrix of two different thermoelectric materials. Therefore, the Seebeck-effect-based heat flux sensors require a durable substrate or thick plate to provide mechanical stability. Hence, the conventional sensors are mainly applicable to flat surfaces and the flexibility is limited. Due to the presence of thick substrates, these sensors have large thermal resistance that disturbs the heat flow. Although the sensibility of the Seebeck-effect-based heat flux sensors is proportional to the number of the thermoelectric material junctions and to the sensor size, the complex structure and low mechanical durability make it difficult to enlarge the sensor size. As an alternative to the Seebeck-effect-based heat flux sensors having these limitations, researchers proposed thin-film-based heat flux sensors driven by ANE [24,39,40]. In the ANE-based sensor, a lateral thermopile structure consisting of alternately arranged and serially connected two different conductor wires with different ANE coefficients is used. Owing to this simple thermopile structure and the symmetry of ANE, the device can be thin, making it easy to reduce the thermal resistance. If magnetic materials showing large ANE can be formed on thin flexible sheets, flexible heat flux sensors can be constructed. These features of the ANE-based heat flux sensors are advantageous over the Seebeck-effect-based heat flux sensors.

Figure 6(a) represents a schematic view of the prototypical ANE-based heat flux sensor having a thermopile structure where the one end of a magnetic wire is electrically connected to the opposite end of an adjacent magnetic wire. In this thermopile structure, when the magnetization (heat flux) is along the width (thickness) direction of the wires, the ANE voltage is generated along the length of the wires and the total output voltage gets multiplied by the number of magnetic wires. However, when we pattern the magnetic film into a wire structure, the strong in-plane demagnetization field reduces the remanent magnetization which causes the reduction of the sensitivity of the device at zero field [24]. For wide applications of heat-flux sensors, the zero-field operation is necessary, and the present SmCo-based amorphous films with strong out of plane anisotropy (see Figure 5(c)) are promising candidate.

For the demonstration, we deposited the 100-nm-thick $Sm_{20}Co_{80}$ amorphous film on a 50-μm-thick flexible polyethylene naphthalate (PEN) sheet. We formed 50 $Sm_{20}Co_{80}$ wires with a width of 50 μm and a length of 10 mm arranged in a parallel configuration at an interval of 150 μm over a region of 10 mm. The wires are then connected in series using 50-μm-thick gold wires to make the total device area of $10 \times 10$ mm$^2$. The total length of the $Sm_{20}Co_{80}$ wires for this configuration reaches 500 mm. To evaluate the performance of the heat flux sensor, we made an experimental configuration shown in Figure 6(b). Here, the ANE-based heat flux sensor and a commercially available Seebeck-effect-based heat flux sensor (Energy Eye, model D0001, Denso corporation) were sandwiched by two Cu blocks. To apply an out-of-plane heat current, the bottom Cu block was heated using a ceramic heater, while the top Cu block acts as a heat bath. After the setup reached thermal equilibrium, the voltage ($V$) between the ends of the thermopile structure was measured with applying an in-plane external magnetic field along the width direction of the wires.

As shown in Figure 6(c), clear ANE signals with a hysteresis behavior was observed in the magnetic field dependence of $V$ and their magnitude increases with increasing the heater power. Figure 6(d) represents $V_{ANE}$ as a function of the heat flux ($j_q$) across the device, where $V_{ANE}$ was obtained by extracting the field-odd component of $V$ at $|\mu_0 H| = 0.2$ T and 0.0 T and $j_q$ was obtained from the output of the commercial heat flux sensor (Figure 6(b)). Using the data, the sensitivity of the ANE-based heat flux sensor, $V_{ANE}/j_q$, was estimated to be $1.12\pm0.02 \times 10^{-7}$ VW$^{-1}$m$^2$ and $0.89\pm0.02 \times 10^{-7}$ VW$^{-1}$m$^2$ in the saturation magnetization ($|\mu_0 H|$ = 0.2 T) and remanent magnetization (0.0 T) states, respectively. The experimentally obtained sensitivity is consistent with the $V_{\text{ANE}}/j_q$ value of $0.94\pm0.16 \times 10^{-7}$ VW$^{-1}$m$^2$, determined considering the measured $S_{\text{ANE}}$ value for the uniform Sm$_{20}$Co$_{80}$ film based on the relation [24]:

$$\frac{V_{\text{ANE}}}{j_q} = \frac{S_{\text{ANE}} L_1 L_2}{\kappa(l_1+l_2)}, \qquad (3)$$

where $L_1 L_2$ is the area of the zigzag pattern (Figure 6(a)), $l_1$ is the width of each magnetic wire, $l_2$ is the interval between the adjacent wires, and $\kappa$ is the thermal conductivity of the magnetic material in the out-of-plane direction for the in-plane magnetization configuration. The obtained sensitivity of the present heat flux sensor at the saturation magnetization state is much larger than that in the prototypical PEN//Fe$_{81}$Al$_{19}$-Au[24] thermopile and comparable to the current best results for the MgO//Co$_2$MnGa-Au[12] and Si/SiO$_2$//Mn$_3$Sn-Au[41] thermopiles. Importantly, the present amorphous-Sm$_{20}$Co$_{80}$-based heat flux sensor has 5.6 times higher sensitivity at zero field as compared to the Co$_2$MnGa-based sensor [12]. The comparable $V_{\text{ANE}}/j_q$ values for the devices on the MgO and PEN substrates suggest that the AEE/ANE performance of the present SmCo-based amorphous films is preserved even on flexible substrates.

Here, we have demonstrated heat flux sensing using a prototypical ANE-based heat flux sensor comprising the amorphous Sm$_{20}$Co$_{80}$ thin film on a flexible substrate. Although till now the sensitivity of this sensor is lower than that of Seebeck-effect-based sensors, ANE has strong advantages in flexibility and low thermal resistance, extending applications of heat flux sensors. In the ANE-based sensors, there is a plenty of room to improve the sensitivity with proper optimization of materials and device structures. As can be seen from Equation (3), the higher sensitivity can be achieved by using materials with larger $S_{\text{ANE}}$, reducing $\kappa$ of ANE materials, increasing the density of the wires (by reducing $l_1$ and $l_2$), replacing Au with a different magnetic material having $S_{\text{ANE}}$ with opposite sign, and increasing the area of the sensor. As shown later, amorphous materials are also beneficial for reducing $\kappa$. Since amorphous materials can be fabricated on any surfaces, they enable large scale manufacturing and reduce the device cost. Therefore, the investigation of amorphous materials with large ANE is one of the most promising approaches towards the development of ANE-based heat flux sensors and our demonstration showcases a guideline for that.

### *3.5. Effect of Fe substitution for Co in $Sm_{20}Co_{80}$ film*

To explore the possibility of further enhancement of ANE/AEE in the $Sm_pCo_{100-p}$ amorphous films, we systematically replaced Co with Fe by fabricating $Sm_{20}(Co_{100-q}Fe_q)_{80}$ composition-spread film with the Fe concentration (*q*) varying from 0 at.% ($Sm_{20}Co_{80}$) to 100 at.% ($Sm_{20}Fe_{80}$) (see Figure 7(a)). Here, $Sm_{20}Co_{80}$ was chosen as the base alloy composition to tune Co concentration due to the presence of the maximum AEE signal and large coercivity. In a similar manner to $Sm_{20}Co_{80}$, the $Sm_{20}(Co_{100-q}Fe_q)_{80}$ alloys are amorphous in nature for the whole composition range, as confirmed by the position-dependent XRD measurement (see Figure 7(b)). The XRD data suggest that the introduction of Fe in $Sm_{20}(Co_{100-q}Fe_q)_{80}$ alloys does not affect their structural configuration. To investigate the microstructures, we performed the cross-sectional STEM analysis. Figures 7(c-d) respectively show the cross-sectional BF-STEM image with the microbeam electron diffraction pattern and STEM-EDS maps of Sm, Co, and Fe obtained from the $Sm_{20}(Co_{77}Fe_{23})_{80}$ composition in the $Sm_{20}(Co_{100-q}Fe_q)_{80}$ composition-spread film. The microbeam electron diffraction pattern clearly confirms the amorphous nature of the film, as determined by XRD. The composition line profiles obtained from a dotted rectangle in STEM-EDS map (Figure 7(d)) are shown in Figure 7(e)**.** Figure 8(a) represents the *q* dependence of in-plane $\mu_0 H_C^{Kerr}$ values for the $Sm_{20}(Co_{100-q}Fe_q)_{80}$ composition-spread film, estimated from the MOKE measurements. The $Sm_{20}(Co_{100-q}Fe_q)_{80}$ alloys show typical ferromagnetic nature with finite coercivity, although the $\mu_0 H_C^{Kerr}$ value was observed to decrease monotonically with increasing the Fe concentration. The origin of such magnetization behavior is yet to be understood and future investigations on microstructural and magnetic evaluations on these films are required.

To investigate the effect of the Fe substitution in the transverse thermoelectric conversion property, we measured the AEE signals for the $Sm_{20}(Co_{100-q}Fe_q)_{80}$ composition-spread film using the LIT technique under the same experimental configuration as described in Section 3.2. Figure 8(b) shows the $A_{AEE}$ and $\phi_{AEE}$ images for the $Sm_{20}(Co_{100-q}Fe_q)_{80}$ composition-spread film in the saturation magnetization ($|\mu_0H|$ = 0.8 T) and remanent magnetization ($|\mu_0H|$ = 0.0 T) states. Figure 8(c) shows the extracted line profile of $A_{AEE}/j_c$ as a function of $q$. These results indicate that the AEE-induced temperature modulation in $Sm_{20}(Co_{100-q}Fe_q)_{80}$ can be enhanced by 20% compared to $Sm_{20}Co_{80}$. The maximum $A_{AEE}/j_c$ value was observed for the $23 \leq q \leq 38$ at.% region. It was found that $A_{AEE}/j_c$ gradually decreases for $q > 38$ at.% and the finite AEE-induced temperature modulation appears for $q > 90$ at.%. The $\phi_{AEE}$ image in Figure 8(b) shows the sign reversal of the AEE signals, i.e., the ~180° change in $\phi_{AEE}$, around $q$ = 90 at.%. We also checked that the AEE signals in Sm-Fe alloys are opposite in sign to those in the Sm-Co alloys (see supplementary material Figure S2). Figure 8(d) shows that the electrical resistivity for the $Sm_{20}(Co_{100-q}Fe_q)_{80}$ composition-spread film is almost independent of $q$. From these results, we determined that the $Sm_{20}(Co_{100-q}Fe_q)_{80}$ alloys with $23 \leq q \leq 38$ at.% show enhanced ANE/AEE. We then quantitatively determine $S_{ANE}$ and $\Pi_{AEE}$ based on the LIT results via the following relations:

$$\Pi_{AEE} = S_{ANE}T, \qquad (4)$$

$$\Pi_{AEE} = \frac{\pi \kappa \Delta T}{4 j_c L}, \qquad (5)$$

where $\Delta T = A_{AEE}\cos\phi_{AEE}$ is the AEE-induced temperature change, and $(4/\pi)j_c$ is the sinusoidal amplitude of the charge current density applied to the magnetic material during the AEE measurement. The out-of-plane $\kappa$ for the $Sm_{20}(Co_{100-q}Fe_q)_{80}$ alloys with different compositions were measured using the TDTR method (see experimental section). As shown in Figure 8(e), the $\kappa$ value for $Sm_{20}(Co_{100-q}Fe_q)_{80}$ is ~6 $Wm^{-1}K^{-1}$ and nearly independent of the Fe concentration, where $\kappa$ of the amorphous alloys is more than 60% smaller than that of the bulk counterpart.[8,11] Therefore, the enhancement in $A_{AEE}/j_c$ directly leads to the increase in $\Pi_{AEE}$. In Figure 8(f), we plot the $\Pi_{AEE}$ and $S_{ANE}$ values for various $Sm_{20}(Co_{100-q}Fe_q)_{80}$ alloys with different Fe concentrations. The highest $S_{ANE}$ of $1.26\pm0.20 \times10^{-6}$ ($0.86\pm0.16 \times10^{-6}$) $VK^{-1}$ was obtained for $Sm_{20}(Co_{77}Fe_{23})_{80}$ at $|\mu_0H| = 0.8$ T ($|\mu_0H| = 0.0$ T).

To confirm the consistency of the result, we fabricated a uniform $Sm_{20}(Co_{77}Fe_{23})_{80}$ film with the optimized composition by alternately depositing Sm, Co, and Fe thin layers (see Figure 9(a)). The structural and magnetic behaviors of the $Sm_{20}(Co_{77}Fe_{23})_{80}$ film were characterized by microbeam electron diffraction patterns (see supplementary material Figure S2) and in-plane magnetization measurement (see Figure 9(b)). The results confirm that the film has similar structural and magnetic behavior as the composition-spread film at the same composition region. $S_{ANE}$ for the uniform $Sm_{20}(Co_{77}Fe_{23})_{80}$ film was directly measured and estimated to be $1.55\pm0.20 \times10^{-6}$ $VK^{-1}$ as shown in Figure 9(c). The directly measured $S_{ANE}$ value for the film is consistent with the value estimated from the AEE measurements using the $Sm_{20}(Co_{100-q}Fe_q)_{80}$ composition-spread film. The maximum $S_{ANE}$ value for the present alloys is still lower than that for the $SmCo_5$-type bulk permanent magnets [8,11], magnetic Heusler alloys (e.g., $Co_2MnGa$ and $Co_2MnSi_{1-x}Al_x$) [15–17,25], and iron-based binary ferromagnets (e.g., $Fe_3Al$ and $Fe_3Ga$) [24,42]. However, as shown before, the unique physical properties of present amorphous alloys, e.g., large in-plane coercivity and remanent magnetization, make them advantageous over the conventional alloys, especially, for zero-field operation of thermoelectric generation and application to heat flux sensors on flexible substates [24].

### 3.6. Detailed characterization of $Sm_{20}Co_{80}$ and $Sm_{20}(Co_{77}Fe_{23})_{80}$ films

Finally, let us discuss the factors determining ANE in our amorphous alloys. $S_{\text{ANE}}$ can be expressed as [23,25],

$$S_{\text{ANE}} = \rho_{xx}\alpha_{xy} + \rho_{xy}\alpha_{xx} \equiv S_{\text{I}} + S_{\text{II}}, \qquad (6)$$

where $\rho_{xx}$ ($\rho_{xy}$) are the diagonal (off-diagonal) component of the electrical resistivity tensor and $\alpha_{xx}$ ($\alpha_{xy}$) are the diagonal (off-diagonal) component of the thermoelectric conductivity tensor. Equation (6) shows that there are two different phenomenological sources in $S_{\text{ANE}}$, and we denote them as $S_{\text{I}} = \rho_{xx}\alpha_{xy}$ and $S_{\text{II}} = \rho_{xy}\alpha_{xx}$ for simplicity. The contribution of $S_{\text{I}}$ is often regarded as an intrinsic part of ANE. The origin of $S_{\text{I}}$ is the direct generation of the transverse electric field originating from the transverse thermoelectric conductivity $\alpha_{xy}$ [43]. The second term, $S_{\text{II}}$, is the contribution of the anomalous Hall effect (AHE) to ANE induced by a longitudinal current due to the Seebeck effect [23,25]. The term can be also expressed as $S_{\text{II}} = \rho_{xy}\alpha_{xx} = S_{\text{SE}}\rho_{xy}/\rho_{xx}$, where $S_{\text{SE}} = \rho_{xx}\alpha_{xx}$. Therefore, by measuring $S_{\text{ANE}}, \rho_{xx}, \rho_{xy}$, and $S_{\text{SE}}$, the $\alpha_{xy}$ value can be estimated experimentally using Equation (6). Figure 10(a) shows the out-of-plane magnetic field dependence of the Hall resistivity $\rho_{\text{Hall}}$ for the $Sm_{20}Co_{80}$ and $Sm_{20}(Co_{77}Fe_{23})_{80}$ films. The Hall resistivity was observed to increase with the field and saturate when the magnetization saturates, indicating the dominant contribution from AHE. The $\rho_{xy}$ values were estimated from the $\rho_{\text{Hall}}$ data by finding the zero-field intercept of the linear-fitted line for the saturation region. We observed that the $\rho_{xy}$ value is enhanced by 114 % for the Fe-substituted alloy (see the inset to Figure 10(a)). Figure 10(b) shows the $\rho_{xx}$ and $S_{\text{SE}}$ values for the uniform $Sm_{20}Co_{80}$ and $Sm_{20}(Co_{77}Fe_{23})_{80}$ films. The anomalous Hall angle ($\theta_{\text{AHE}} = -\rho_{xy}/\rho_{xx}$) and $\sigma_{xy}$ ($= -\frac{\rho_{xy}}{\rho_{xx}^2+\rho_{xy}^2}$) were estimated from the $\rho_{xx}$ and $\rho_{xy}$ data. Figure 10(c) shows a huge enhancement in the $\theta_{\text{AHE}}$ and $\sigma_{xy}$ values for $Sm_{20}(Co_{77}Fe_{23})_{80}$ as compared to $Sm_{20}Co_{80}$. Using the experimentally observed values, we estimated the $S_{\text{I}}$, $S_{\text{II}}$, and $\alpha_{xy}$ values using Equation (6). Figure 10(d) shows that, for both the $Sm_{20}Co_{80}$ and $Sm_{20}(Co_{77}Fe_{23})_{80}$ alloys, the major contribution to ANE comes from the $S_I$ term and the $S_{II}$ term is very small due to small $S_{SE}$. We found that the enhancement in $S_{ANE}$ for the $Sm_{20}(Co_{77}Fe_{23})_{80}$ alloy is attributed to the $S_I$ enhancement. Although $\alpha_{xy}$ for $Sm_{20}(Co_{77}Fe_{23})_{80}$ is 73% larger than that for $Sm_{20}Co_{80}$, the value is an order of magnitude smaller than that for the $SmCo_5$-type bulk permanent magnets [8,11], resulting in smaller $S_{ANE}$ in our amorphous films. Further improvement of $S_{ANE}$ and $\alpha_{xy}$ in magnetic amorphous films is one of the future tasks.

## 4. Conclusions

In this study, we have demonstrated that Sm-Co-based amorphous alloy films exhibit large in-plane remanent magnetization and coercive force and enable zero-field operation of transverse thermoelectric generation, i.e., ANE and AEE. First, we systematically investigated the composition dependence of AEE for the $Sm_pCo_{100-p}$ composition-spread film with $p$ varying from 0 to 100 at.% by the combinatorial sputtering technique. The experiments revealed that $Sm_{20}Co_{80}$ exhibits the largest AEE-induced temperature modulation among the $Sm_pCo_{100-p}$ amorphous alloys. By directly measuring ANE, the ANE coefficient of the $Sm_{20}Co_{80}$ amorphous alloy was estimated to be $1.07\pm0.08 \times 10^{-6}$ VK$^{-1}$. To explore the usability of our films, we constructed a prototypical ANE-based heat flux sensing device using the $Sm_{20}Co_{80}$ amorphous film fabricated on a flexible PEN substrate. The obtained sensitivity for this prototypical ANE-based thermopile at zero filed is comparable to the previously reported best results on ANE-based thermopiles. The demonstration opens the door to exploring amorphous ANE/AEE materials for flexible thermoelectric applications.

To explore the possibility of further enhancement of ANE/AEE for the Sm-Co-based amorphous films, we systematically replaced Co with Fe by fabricating the $Sm_{20}(Co_{100-q}Fe_q)_{80}$ composition-spread film with $q$ varying from 0 to 100 at.%. Significant enhancement in the AEE-induced temperature modulation was observed in the alloy with 23 at.% Fe substitution. The ANE coefficient of the $Sm_{20}(Co_{77}Fe_{23})_{80}$ amorphous film was estimated to be $1.55\pm0.20 \times10^{-6}$ VK$^{-1}$. Despite of the smaller $S_{ANE}$ values, the presence of large in-plane remanent magnetization and coercive force in our films makes it advantageous over conventional materials for applications including ANE-based heat flux sensors, as these materials can generate thermoelectric output without an external magnetic field. Importantly, owing to the amorphous nature, the films can be fabricated on any surfaces, increasing the versatility of ANE/AEE devices. Exploring the possibility of hybrid films consisting of the present films and other magnetic materials will be important to realize the zero-field operation of large ANE/AEE.

**Declaration of Competing Interest**

The authors declare that they have no known competing financial interests or personal relationships that could have appeared to influence the work reported in this paper.

**Acknowledgements**

The authors thank R. Iguchi for valuable discussions and I. Narita for technical support to do the composition analysis. This work was supported by CREST "Creation of Innovative Core Technologies for Nano-enabled Thermal Management" (JPMJCR17I1 and JPMJCR17I2) and PRESTO "Scientific Innovation for Energy Harvesting Technology" (JPMJPR17R5) from Japan Science and Technology Agency, Japan; Grant-in-Aid for Scientific Research (S) (18H05246) from Japan Society for the Promotion of Science (JSPS) KAKENHI, Japan; "Mitou challenge 2050" (P14004) from NEDO, Japan; and the NEC Corporation. R.M. is supported by JSPS through the "JSPS Postdoctoral Fellowship for Research in Japan (Standard)" (P21064).

**Data Availability Statement**

The data that support the findings of this study are available from the corresponding authors upon reasonable request.

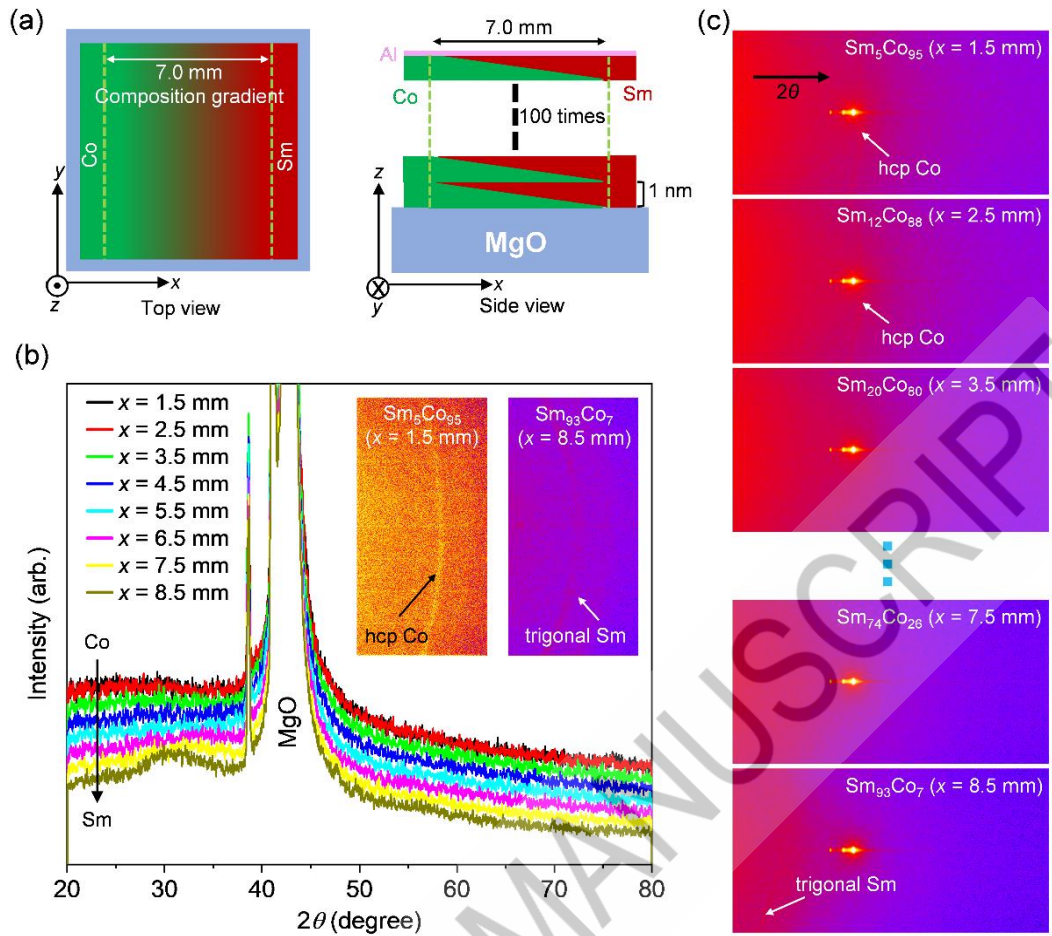

**Figure 1.** (a) Schematic of the $Sm_pCo_{100-p}$ composition-spread film structure, where the wedge-shaped Sm and Co layers were alternatively deposited using the linear moving shutter on a MgO substrate. (b) XRD profiles for the $Sm_pCo_{100-p}$ composition-spread film measured at 1-mm interval along the $x$ direction. (c) Corresponding 2D XRD profile. The compositions written on the 2D profile are estimated from the $x$-dependent composition data measured by thermal emission EPMA. Here $x = 0$ corresponds to the substrate edge at pure Co side.

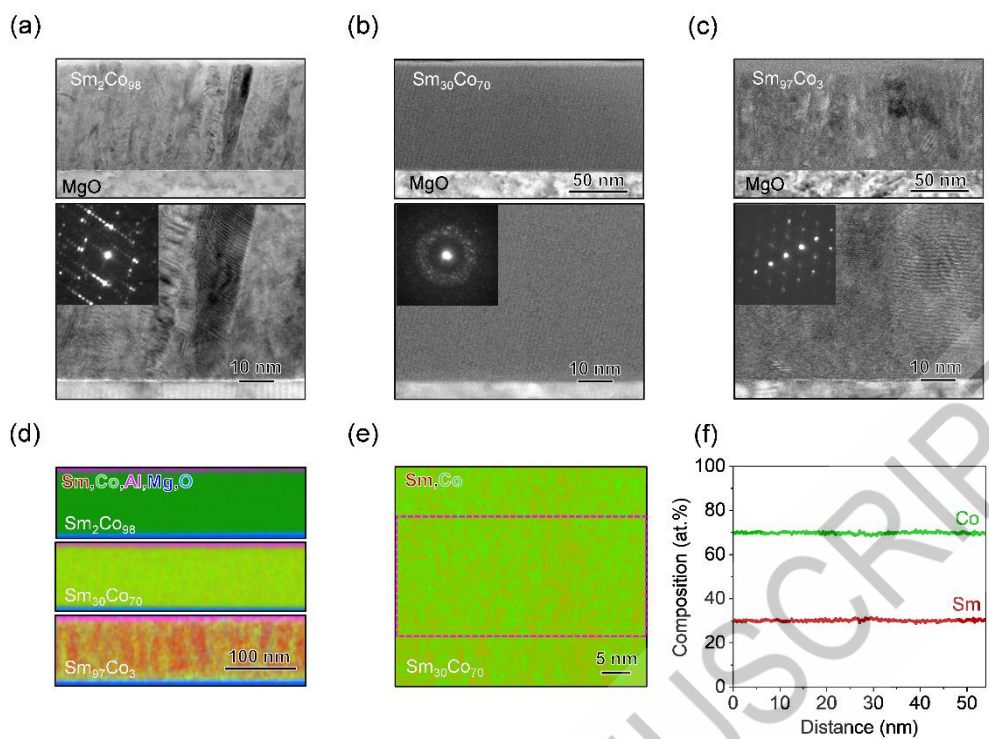

**Figure 2.** (a)-(c) Cross-sectional BF-STEM images and microbeam electron diffraction patterns for the $Sm_pCo_{100-p}$ composition-spread film measured at three different positions corresponding to the compositions (a) $Sm_2Co_{98}$, (b) $Sm_{30}Co_{70}$, and (c) $Sm_{97}Co_3$. (d) The STEM-EDS elemental maps obtained at the same positions of (a)-(c). (e) High magnification STEM-EDS maps of Sm and Co observed from $Sm_{30}Co_{70}$ composition. (f) The composition line profiles of the constituent elements along the horizontal direction obtained from the selected box region shown in (e).

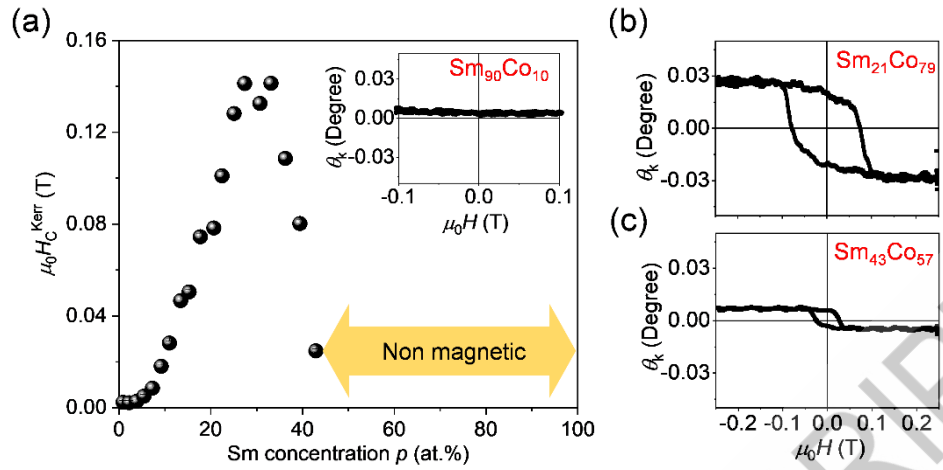

**Figure 3.** (a) Sm concentration $p$ dependence of the coercive field $\mu_0 H_C^{\text{Kerr}}$ estimated from the MOKE loops for the $\text{Sm}_p\text{Co}_{100-p}$ composition-spread film, measured with applying an in-plane magnetic field. The inset shows the absence of the Kerr loop at a non-magnetic Sm-rich region of the $\text{Sm}_p\text{Co}_{100-p}$ composition-spread film. (b),(c) In-plane Kerr loops observed for the $\text{Sm}_p\text{Co}_{100-p}$ composition-spread film at the $\text{Sm}_{21}\text{Co}_{79}$ and $\text{Sm}_{43}\text{Co}_{54}$ compositions.

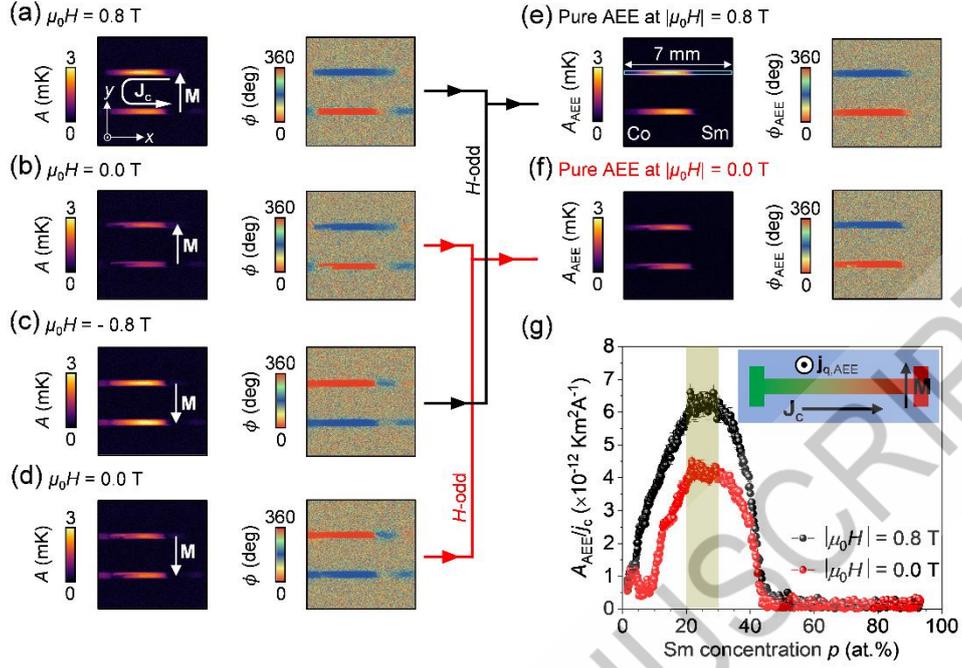

**Figure 4.** (a)-(d) Lock-in amplitude $A$ and phase $\phi$ images for the $Sm_pCo_{100-p}$ composition-spread film recorded by applying a square-wave-modulated AC charge current with the amplitude $J_c = 15$ mA and frequency $f = 25$ Hz for various values of the in-plane external magnetic field $\mu_0 H$. $\mathbf{J}_c$ and $\mathbf{M}$ represent the charge current and magnetization directions, respectively. (e) $A_{AEE}$ and $\phi_{AEE}$ images showing the pure AEE-induced temperature modulation for the $Sm_pCo_{100-p}$ composition-spread film at $|\mu_0 H| = 0.8$ T, obtained by subtracting the images in (c) from those in (a). (f) $A_{AEE}$ and $\phi_{AEE}$ images at $|\mu_0 H| = 0.0$ T, obtained by subtracting the images in (d) from those in (b). (g) Line profile of $A_{AEE}$ normalized by the charge current density $j_c$ at $|\mu_0 H| = 0.8$ T and 0.0 T as a function of the Sm concentration $p$.

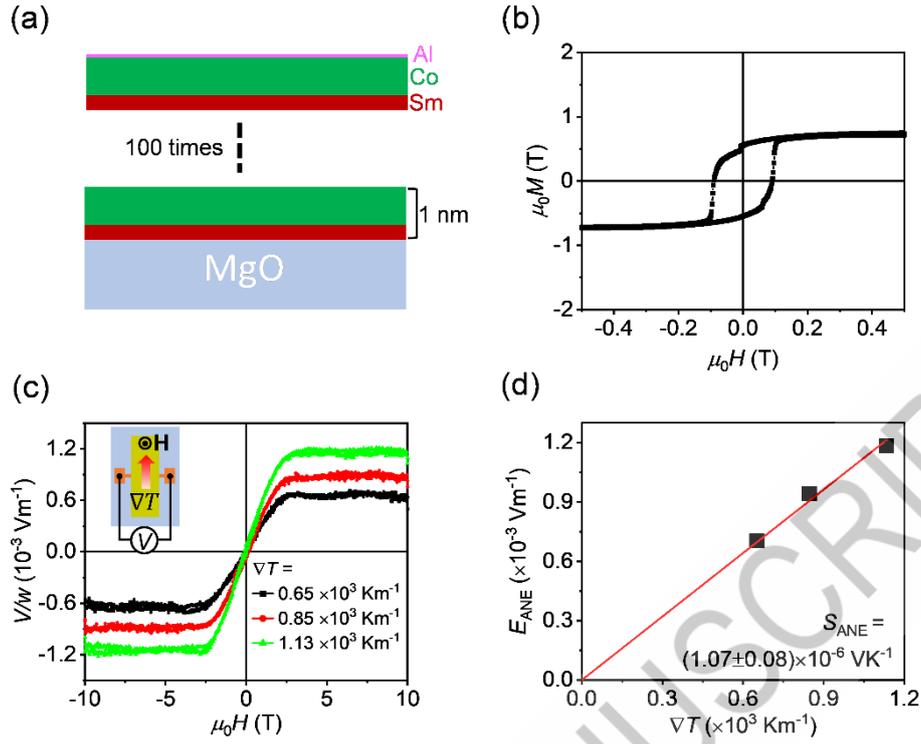

**Figure 5.** (a) Schematic of the uniform $Sm_{20}Co_{80}$ film structure, where the Sm/Co layers were alternatively deposited on a MgO substrate. The thickness of each layer was optimized from the composition-spread film to get the desired compositions. The composition of the deposited uniform films was verified by EPMA measurements. (b) $H$ dependence of the magnetization $M$ for the film at room temperature. (c) $H$ dependence of the transverse voltage $V$ normalized by the width of the Hall bar $w$ for various values of the temperature gradient $\nabla T$. (d) $\nabla T$ dependence of the ANE-induced electric field $E_{ANE}$ for the uniform $Sm_{20}Co_{80}$ film, extracted by the field-odd component of the voltage at $\mu_0 H = 5.0$ T.

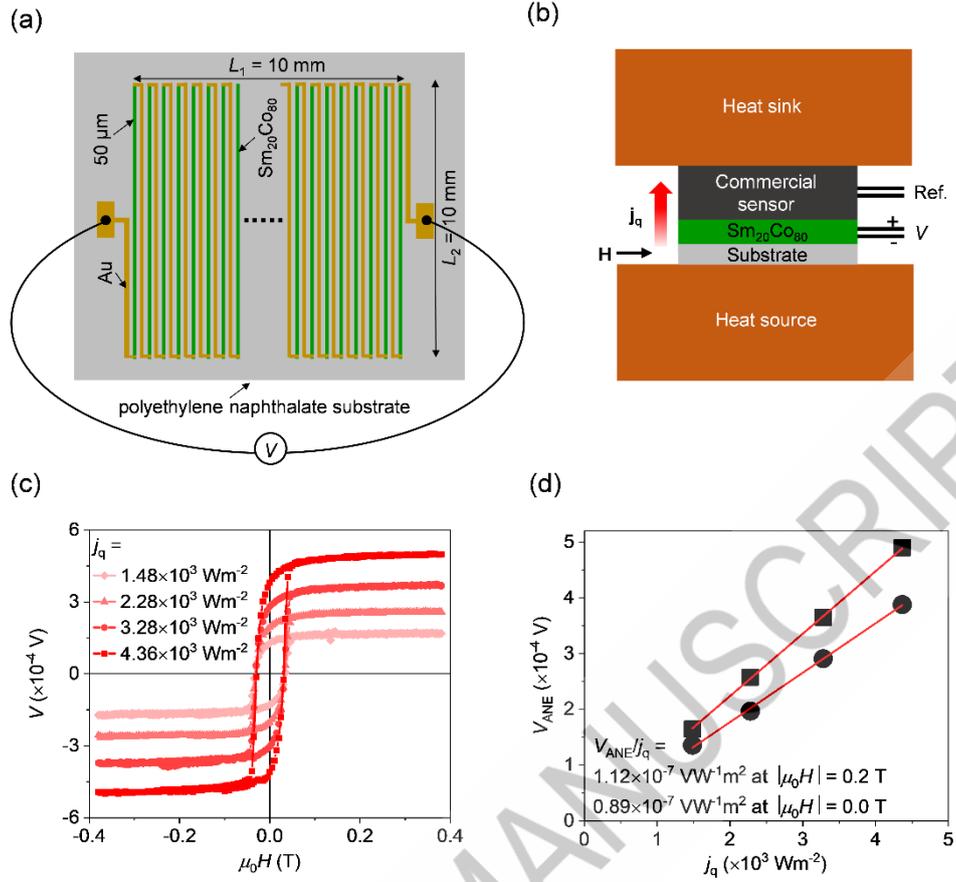

**Figure 6.** (a) Schematic of the ANE-based heat flux sensor. The sensor consists of the $Sm_{20}Co_{80}$ and Au wires arranged alternately and connected in series on a flexible polyethylene naphthalate substrate. (b) Schematic of the experimental setup to evaluate the sensitivity of the ANE-based heat flux sensor. (c) $H$ dependence of $V$ from the $Sm_{20}Co_{80}$/Au-based sensor for various values of the heat flow $j_q$. (d) $j_q$ dependence of $V_{ANE}$ at $|\mu_0 H| = 0.2$ and 0.0 T. $V_{ANE}$ is extracted by the field-odd component of $V$ at $\mu_0 H = 0.2$ T. $V_{ANE}$ at zero field was extracted from the zero field $V$ values for the positive-to-negative and negative-to-positive sweep of $H$. The solid lines show the results of the linear fitting.

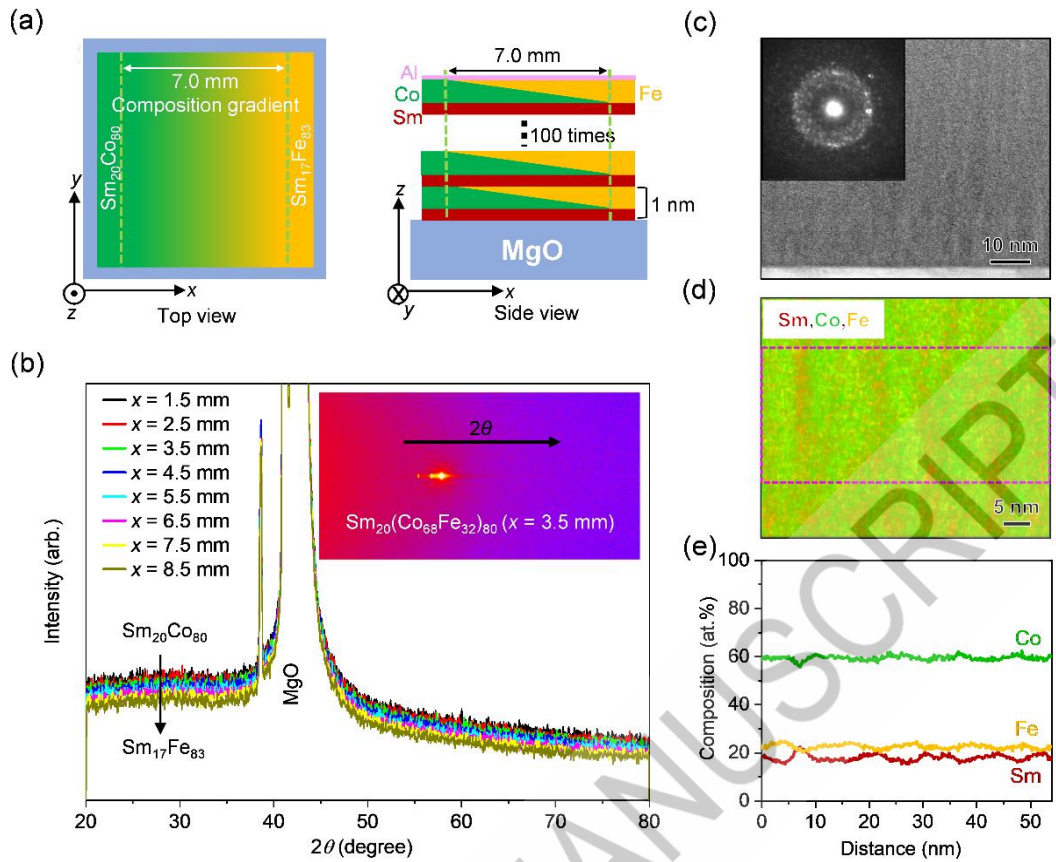

**Figure 7.** Schematic of the $Sm_{20}(Co_{100-q}Fe_q)_{80}$ composition-spread film structure, where the uniform Sm layer and wedge-shaped Co and Fe layers were alternatively deposited using the linear moving shutter on a MgO substrate. (b) XRD profiles for the $Sm_{20}(Co_{100-q}Fe_q)_{80}$ composition-spread film measured at the 1-mm interval along the $x$ direction. The inset shows 2D XRD profile at $x = 3.5$ mm. Here $x = 0$ corresponds to the substrate edge at pure $Sm_{20}Co_{80}$ side. (c) Cross-sectional BF-STEM image and microbeam electron diffraction pattern for the $Sm_{20}(Co_{100-q}Fe_q)_{80}$ composition-spread film measured near the $Sm_{20}(Co_{77}Fe_{23})_{80}$ composition. (d) High magnification STEM-EDS maps for the $Sm_{20}(Co_{77}Fe_{23})_{80}$ composition. (e) The composition line profiles of the constituent elements along the horizontal direction obtained from the dotted rectangle region shown in (d).

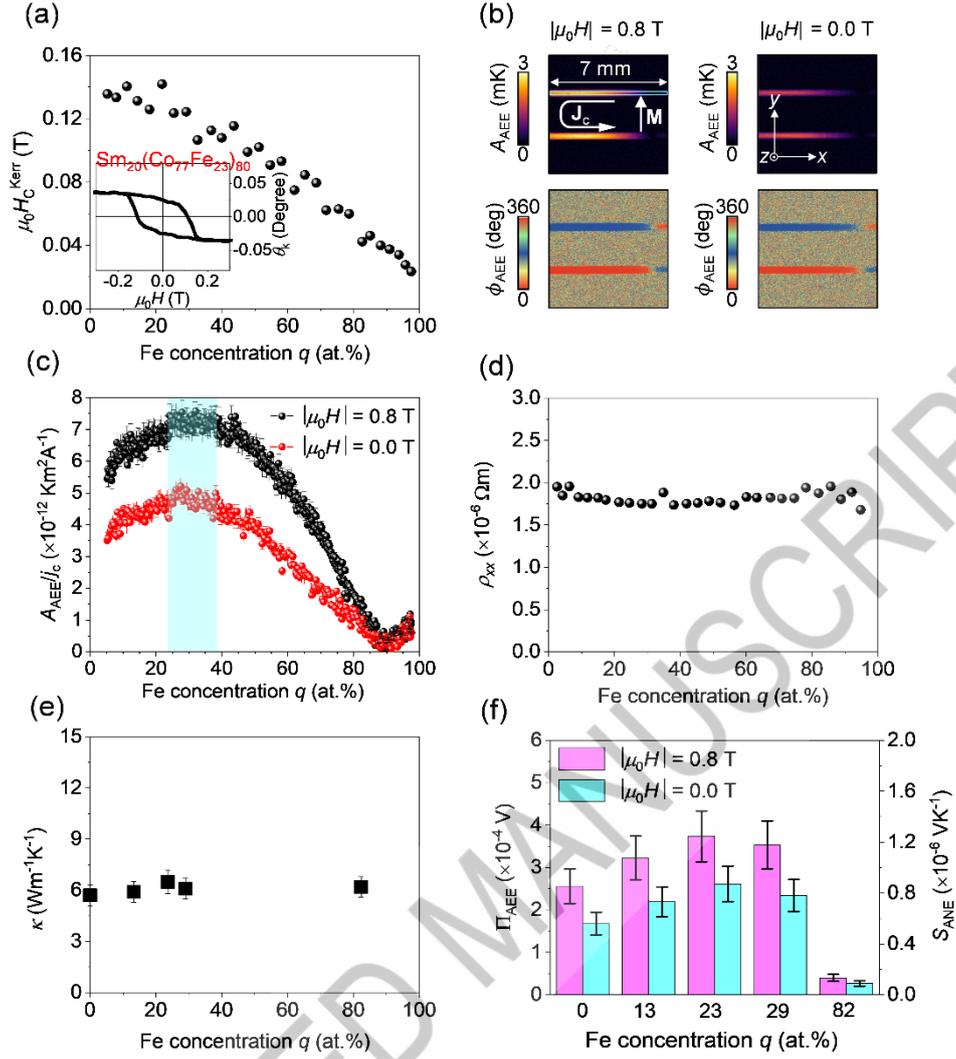

**Figure 8.** (a) Fe concentration $q$ dependence of the coercive field $\mu_0 H_C^{\text{Kerr}}$ estimated from the MOKE loops for the $\text{Sm}_{20}(\text{Co}_{100-q}\text{Fe}_q)_{80}$ composition-spread film, measured with applying an in-plane magnetic field. (b) $A_{\text{AEE}}$ and $\phi_{\text{AEE}}$ images showing the pure AEE-induced temperature modulation for the $\text{Sm}_{20}(\text{Co}_{100-q}\text{Fe}_q)_{80}$ composition-spread film at $J_c = 15$ mA, $f = 25$ Hz, and $|\mu_0 H| = 0.8$ T and 0.0 T. (c) Line profile of $A_{\text{AEE}}/j_c$ at $|\mu_0 H| = 0.8$ T and 0.0 T as a function of $q$. (d) $q$ dependence of the electrical resistivity ($\rho_{xx}$) measured the using $\text{Sm}_{20}(\text{Co}_{100-q}\text{Fe}_q)_{80}$ composition-spread film. (e) Out-of-plane thermal conductivity $\kappa$ measured by the time-domain thermoreflectance method using the $\text{Sm}_{20}(\text{Co}_{100-q}\text{Fe}_q)_{80}$ composition-spread film. (f) ANE coefficient $S_{\text{ANE}}$ and AEE coefficient $\Pi_{\text{AEE}}$ for the $Sm_{20}(Co_{100-q}Fe_q)_{80}$ alloys at different values of $q$, estimated from the AEE data using Equation (4) and (5).

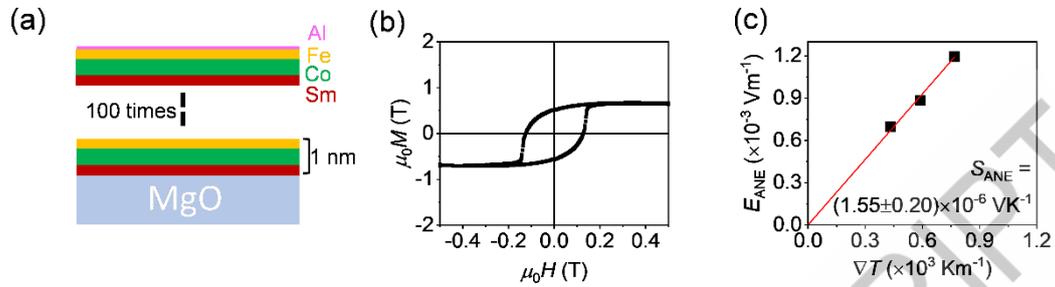

**Figure 9.** (a) Schematics of the uniform $Sm_{20}(Co_{77}Fe_{23})_{80}$ film structure, where the Sm, Co, and Fe layers were alternatively deposited on a MgO substrate. The composition of the deposited uniform films was verified by EPMA measurements. (b) $H$ dependence of $M$ for the film at room temperature. (c) $\nabla T$ dependence of $E_{ANE}$ for the uniform $Sm_{20}(Co_{77}Fe_{23})_{80}$ film, extracted by the field-odd component of the voltage at $|\mu_0 H| = 5.0$ T.

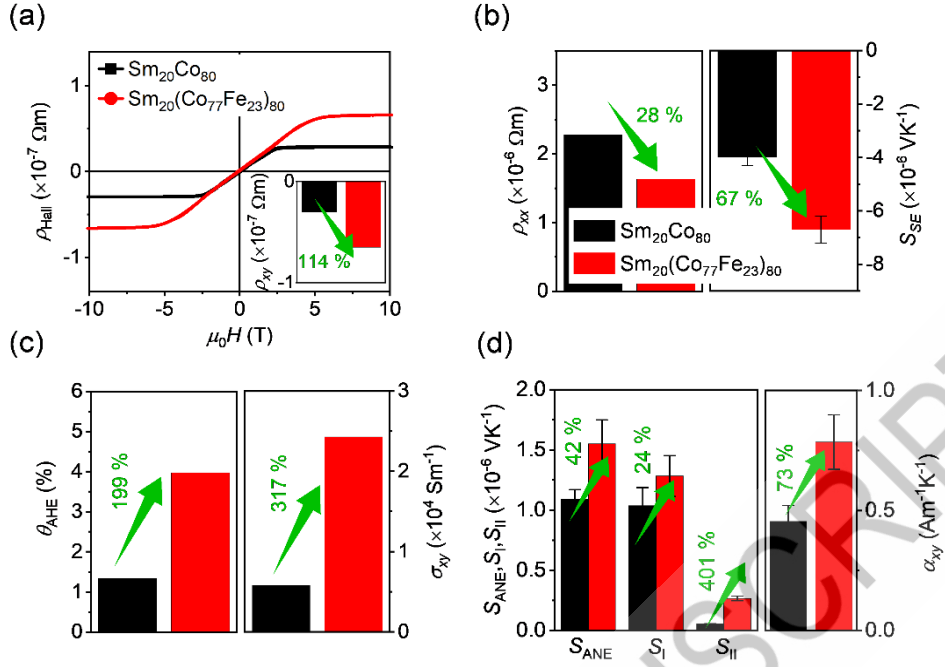

**Figure 10.** (a) $H$ dependence of the Hall resistivity $\rho_{Hall}$ for the uniform $Sm_{20}Co_{80}$ and $Sm_{20}(Co_{77}Fe_{23})_{80}$ films. The inset shows the anomalous Hall resistivity $\rho_{xy}$ estimated from the $\rho_{Hall}$ data by extracting the zero-field intercept of the linear-fitted line for the saturation region. (b) Electrical resistivity $\rho_{xx}$ and Seebeck coefficient $S_{SE}$ for the uniform $Sm_{20}Co_{80}$ and $Sm_{20}(Co_{77}Fe_{23})_{80}$ films. (c) The anomalous Hall angle $\theta_{AHE}$ and anomalous Hall conductivity $\sigma_{xy}$. (d) Extracted $S_{ANE}$, $S_{I}$, $S_{II}$, and transverse thermoelectric conductivity $\alpha_{xy}$ for the uniform $Sm_{20}Co_{80}$ and $Sm_{20}(Co_{77}Fe_{23})_{80}$ films.

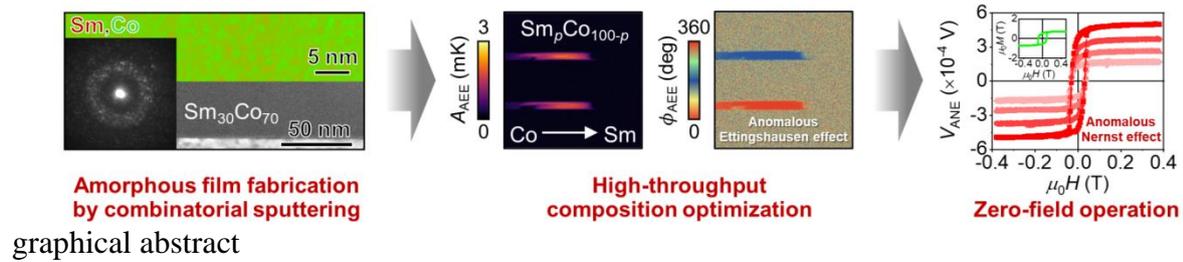
graphical abstract

# Supplementary material

# Sm-Co-based amorphous alloy films for zero-field operation of transverse thermoelectric generation

Rajkumar Modak[a], Yuya Sakuraba[a], Takamasa Hirai[a], Takashi Yagi[b], Hossein Sepehri-Amin[a], Weinan Zhou[a], Hiroto Masuda[c], Takeshi Seki[a,c,d], Koki Takanashi[c,d,e,*], Tadakatsu Ohkubo[a], and Ken-ichi Uchida[a,c,d]

[a]National Institute for Materials Science, Tsukuba, 305-0047, Japan

[b]National Institute of Advanced Industrial Science and Technology, Tsukuba 305-8565, Japan

[c]Institute for Materials Research, Tohoku University, Sendai 980-8577, Japan

[d]Center for Spintronics Research Network, Tohoku University, Sendai 980-8577, Japan

[e]Center for Science and Innovation in Spintronics, Tohoku University, Sendai 980-8577, Japan

*Present address: Advanced Science Research Center, Japan Atomic Energy Agency, Tokai, Ibaraki 319-1195, Japan

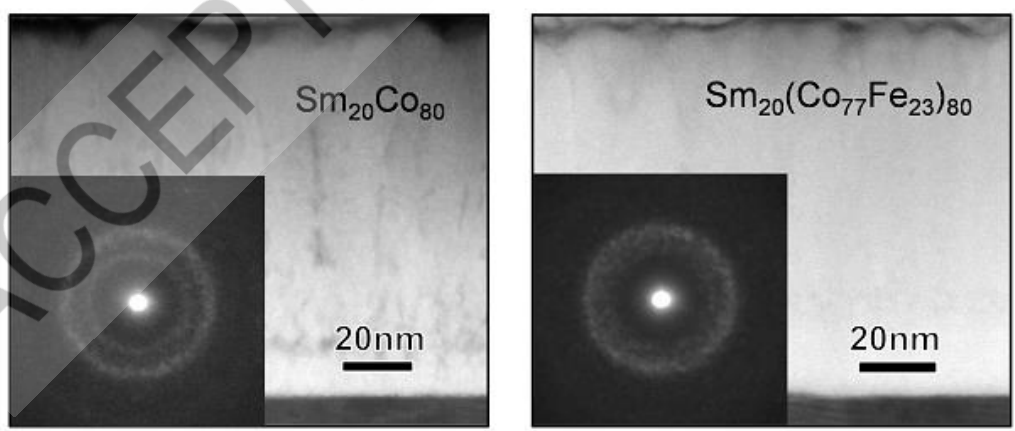

**Figure S1.** Cross-sectional high angle annular dark field (HAADF) scanning transmission electron microscope (STEM) images and microbeam electron diffraction pattern obtained from the uniform $Sm_{20}Co_{80}$ (left) and $Sm_{20}(Co_{77}Fe_{23})_{80}$ (right) films.

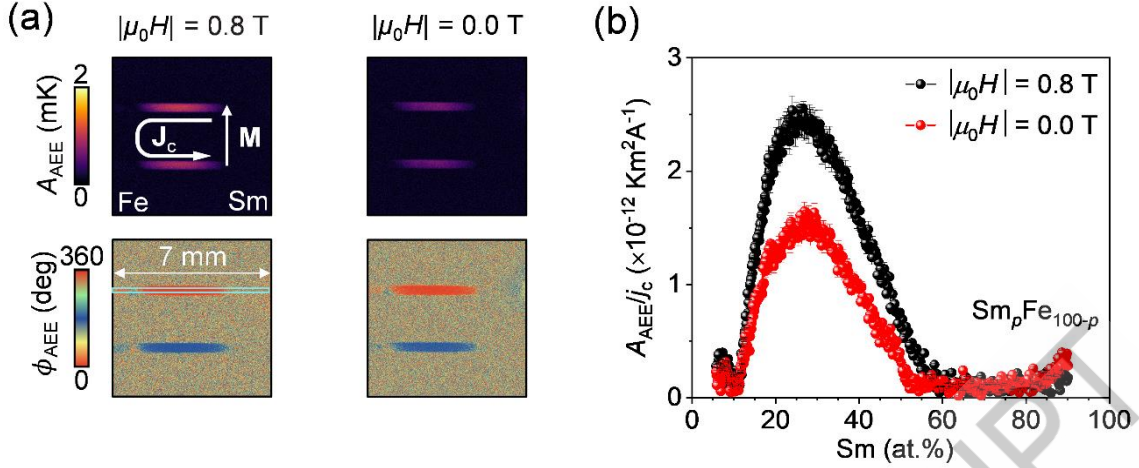

**Figure S2.** (a) Lock-in amplitude $A_{AEE}$ and phase $\phi_{AEE}$ images showing the temperature modulation induced by the anomalous Ettingshausen effect for the charge current amplitude $J_c$ = 15 mA and frequency $f$ = 25 Hz for the $Sm_pFe_{100-p}$ composition-spread film at the in-plane external magnetic fields $|\mu_0 H|$ = 0.8 T and $|\mu_0 H|$ = 0.0 T. (c) Line profile of $A_{AEE}$ normalized by the charge current density $j_c$ at $|\mu_0 H|$ = 0.8 T and $|\mu_0 H|$ = 0.0 T as a function of the Sm concentration $p$.

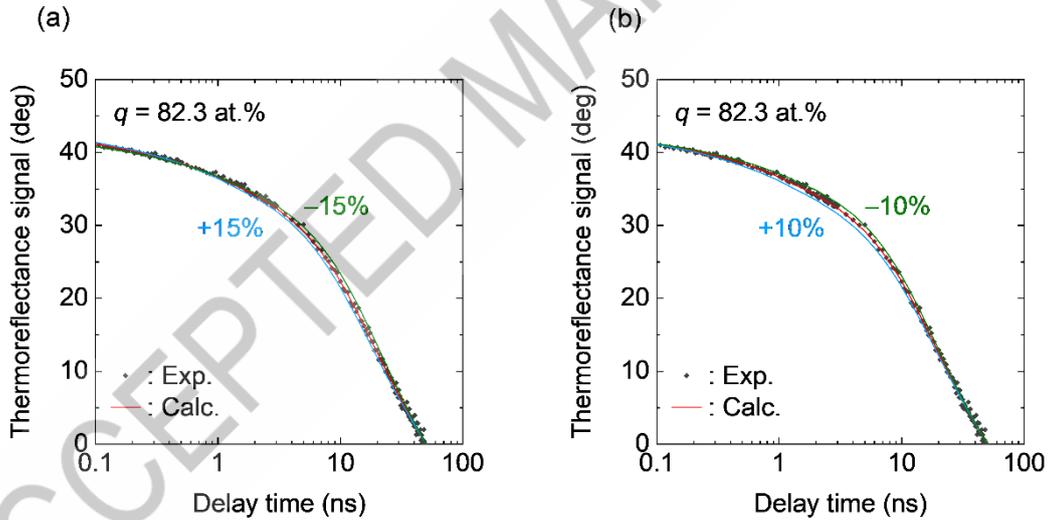

**Figure S3.** (a) Thermal diffusivity and (b) volumetric heat capacity in the TDTR analysis. The figures show the measured TDTR signal (black squares) along with the numerically fitted data (red solid curves) to the heat equation. The green and blue solid curves represent fitting with ±15% and ±10% deviation in the thermal diffusivity and volumetric heat capacity. The magnitude of the error bars for the thermal conductivity in Figure 8(e) was determined by considering the relation between the fitting accuracy and parameters.